\documentclass[structabstract]{aa}  
\usepackage{graphicx}
\usepackage{txfonts}
\usepackage{url}
\usepackage{afterpage}
\usepackage[authoryear]{natbib}
\usepackage{longtable}
\bibpunct{(}{)}{;}{a}{}{,} 
\def\kms{$\mathrm {km~s}^{-1}$}

\begin{document}
   \title{Three carbon-enhanced metal-poor dwarf stars from the SDSS} 

   \subtitle{Chemical abundances from CO$^5$BOLD 3D hydrodynamical model atmospheres
\thanks{Based on observations obtained with the ESO Very Large
Telescope at Paranal Observatory, Chile (programmes 078.D-0217 and 383.D-0927)}}

   \author{
N.T. Behara
\inst{1,2}
\and
P. Bonifacio
\inst{1,2,3}
\and
H.-G. Ludwig 
\inst{1,2}
\and
L. Sbordone
\inst{1,2}
\and
J.I. Gonz\'alez Hern\'andez
\inst{1,2,4}
\and
E. Caffau
\inst{2}
}

   \offprints{N.T. Behara}

   \institute{
CIFIST Marie Curie Excellence Team
         \and
GEPI, Observatoire de Paris
CNRS, Universit\'e Paris Diderot; Place
Jules Janssen 92190
Meudon, France 
\and
Istituto Nazionale di Astrofisica - Osservatorio Astronomico di
Trieste, Via Tiepolo 11, I-34143  Trieste, Italy
\and
Dpto. de Astrof{\'\i}ısica y Ciencias de la Atm\'osfera, Facultad de
Ciencias F{\'\i}sicas, \\
Universidad Complutense de Madrid, E-28040
Madrid, Spain 
    }
\authorrunning{Behara et al.}
\titlerunning{Spectral analyses of three CEMP stars}

   \date{Received ; accepted }

 
  \abstract
   {The origin of carbon-enhanced metal-poor stars enriched with both $s$ and
     $r$ elements is highly debated. Detailed abundances of these types of
     stars are crucial to understand the nature of their progenitors. }
   {The aim of this investigation is to study in detail the abundances of SDSS
     J1349-0229, SDSS J0912+0216 and SDSS J1036+1212, three dwarf CEMP stars, selected 
from the Sloan Digital Sky Survey.}
   {Using high resolution VLT/UVES spectra (R $\sim 30\,000$) we determine
     abundances for Li, C, N, O, Na, Mg, Al, Ca, Sc, Ti, Cr, Mn, Fe, Co, Ni
     and 21 neutron-capture elements. We made use of CO$^5$BOLD 3D
     hydrodynamical model atmospheres in the analysis of the carbon, nitrogen
     and oxygen abundances. NLTE corrections for C{\sc i} and O{\sc i} lines
     were computed using the Kiel code. }
   {We classify SDSS J1349-0229 and SDSS J0912+0216 as CEMP-r+s stars. SDSS
     J1036+1212 belongs to the class CEMP-no/s, with enhanced Ba, but
     deficient Sr, of which it is the third member discovered to
     date. Radial-velocity variations have been observed in SDSS J1349-0229,
     providing evidence that it is a member of a binary system.}
   {The chemical composition of the three stars is generally compatible with 
mass transfer from an AGB companion. However, many details  remain difficult
to explain. Most notably of those are the abundance of Li at the level of the Spite
plateau in SDSS J1036+1212 and the large over-abundance of the
pure r-process element Eu in all three stars. }


   \maketitle
%

\section{Introduction\label{intro}}

Carbon-enhanced metal-poor (hereafter CEMP) stars owe
their name to a considerable overabundance of carbon with
respect to iron ([C/Fe] $>$ +1.0), and represent
a sizeable fraction
of the very metal-poor stars ([Fe/H] $<$ --2.0).
The frequency of CEMP stars 
has been estimated to be of the order of
14\% by \citet{cohen05} and of 
21\% by \citet{lucatello06}.
This frequency
appears to increase with decreasing 
metallicity, reaching approximately 
40\% for stars with [Fe/H] $<$ --3.5 \citep{BC2005}.  
The carbon excess in CEMP stars
may stem from carbon production by nucleosynthesis in an
asymptotic giant branch (AGB) star followed by mass transfer to a
surviving companion. 
An alternative mechanism is the formation of the
star from a C-enriched interstellar medium.

The chemical composition of these stars shows a considerable
variety. 
On the basis of the abundance pattern of neutron capture elements
\citet{BC2005} suggested to divide CEMP stars into 
four classes: CEMP-no, CEMP-s, CEMP-r/s, CEMP-r.
To these classes \citet{sivarani} proposed to add  
the class of CEMP-no/s to accommodate two stars
discovered in the course of the ``First Stars''
programme \citep[see][and references therein for
a description of the programme]{bonifacioFS}.

The CEMP-no stars do not show any enhancement of neutron-capture elements.
The prototype of this class is CS 22957-027 \citep{norris1997,bonifacio1998}.
The class also includes the three most Fe-poor
stars known: HE 0107-5240, HE 1327-2326 and HE 0557-4840 (Christlieb et
al.~2002; Frebel et al.~2005; Norris et al.~2007).

The majority of CEMP stars with [Fe/H] $> -3$ are strongly enhanced in
$s$-process elements (Aoki et al.~2007) and form the class
CEMP-s. Radial-velocity studies by Lucatello et al.~(2005) have shown that  
these may all be members of binary systems, providing strong evidence for the
mass transfer scenario from a more massive companion. The large scatter
of observed $s$-element abundances may arise from the different masses of the
companion stars, which presents a useful constraint for AGB nucleosynthesis models.

Among the stars with strong enhancement of
the $s$-process elements, a high percentage also show large enhancements of
$r$-process elements. These stars belong to the class CEMP-r/s, 
the prototypes of which (CS 22948-27 and CS 29497-34) were 
first studied  by \citet{barbuy} and \citet{hill}.
Many scenarios have been proposed
to explain the observed abundances. Most involve two independant processes, one for the 
$r$-elements and one for the $s$-elements.
None of the scenarios,
however, is entirely convincing given the many different
abundance patterns observed in r/s stars
as well as the uncertainties regarding the astrophysical
sites of the r-process. For example, Johnson \& Bolte
(2004) find r-element ratios which are difficult to explain with
combinations of the normal r and s-processes.

Only one r-process enhanced CEMP star (CEMP-r) has been detected: CS 22892-052
(Sneden et al.~2003). This stars exhibits the lowest carbon enhancement of all
CEMP stars: [C/Fe] = +1.0. The $r$-process pattern observed in the heavier
n-capture elements was found to be in excellent agreement with the scaled
solar $r$-process abundances. Lighter elements with 40 $\leq$ Z $\leq$ 50
however were not found to match, leading the authors to suggest the existence
of two $r$-process sites.

Finally the class of CEMP-no/s was up to now defined
by only two members, CS 29528-041 and CS 31080-095,
which show an enhancement of Ba over iron of about
1\,dex, but a sizeable underabundance of Sr.

The diversity of $s$ and $r$-element patterns among CEMP stars strongly suggests
that this population of stars has several astrophysical origins. 
Other mechanisms may be responsible for their nucleosynthetic history
beyond those involving AGB stars.
Since the $s$-process and $r$-process occur under
different physical conditions, they are likely to arise in different
astrophysical sites. 

In this paper we present detailed abundance analyses of three dwarf CEMP
stars: SDSS J1349-0229, SDSS J0912+0216 and SDSS J1036+1212. These stars are
extremely metal-poor, with [Fe/H] $<$ --2.50, and were selected from our
ongoing survey of extremely metal-poor dwarf candidates from the Sloan Digital
Sky Survey. The first two belong to the CEMP-r+s class, while
the third is the third known example of a CEMP-no/s star, after the
first two discovered by \citet{sivarani}, suggesting that the
class is probably not so rare. 

\begin{table*}
\caption{Log of observations. }
\begin{tabular}[b!]{lccccccc}
\hline\hline
Star            & $\alpha$ ~~~~~~~~~~~~ $\delta$ & Date \& Time         & Exposure  & Setting &  Heliocentric& $g$ \\
                &                                &                      &  time     &         &  radial velocity                &    \\
                &  (J2000)                       & yyyy-mm-dd~ UT       &  (s)      &         &  \kms            & mag \\
\hline
\\
SDSS J1349-0229 & 13 49 13.5 --02 29 42.8 &  2007-01-31 07:32:54 & 3800  & Dic1 390+580  &92.3 & 16.63 \\
                &                        &  2009-04-15 02:24:25 & 3005  & Dic2 346+760  & 121.8 & 16.63 \\
                &                        &  2009-04-15 03:54:58 & 3005  & Dic2 346+760  & 121.7 & 16.63 \\
                &                        &  2009-04-15 04:52:57 & 3005  & Dic2 346+760  & 122.0 & 16.63 \\
                &                        &  2009-04-15 05:45:07 & 3005  & Dic2 346+760  & 122.4 & 16.63 \\
SDSS J0912+0216 & 09 12 43.7 +02 16 23.7  &  2006-12-31 07:41:05 & 3005  & Dic1 390+580  &138.0 & 15.67\\
SDSS J1036+1212 & 10 36 49.9 +12 12 19.8  &  2007-01-29 07:16:27 & 3005  & Dic1 390+580  &--47.9 & 16.59\\
\hline
\end{tabular}
\label{tab_obs}
\end{table*}

\section{Target selection and observations}

We selected extremely metal-poor (EMP) dwarf candidates from the Sloan Digital Sky Survey 
\citep[SDSS;][]{york,dr5,dr6} with an automatic analysis code \citep{BC03}
in a version which provides estimates of [M/H]
from low resolution spectra. These candidates were observed 
at high resolution  with
the UVES spectrograph \citep{dekker} at the ESO 8.2m Kueyen-VLT telescope
as part of our survey of stars at low metallicity.

\begin{figure*}
\centering \includegraphics[width=1.0\textwidth]{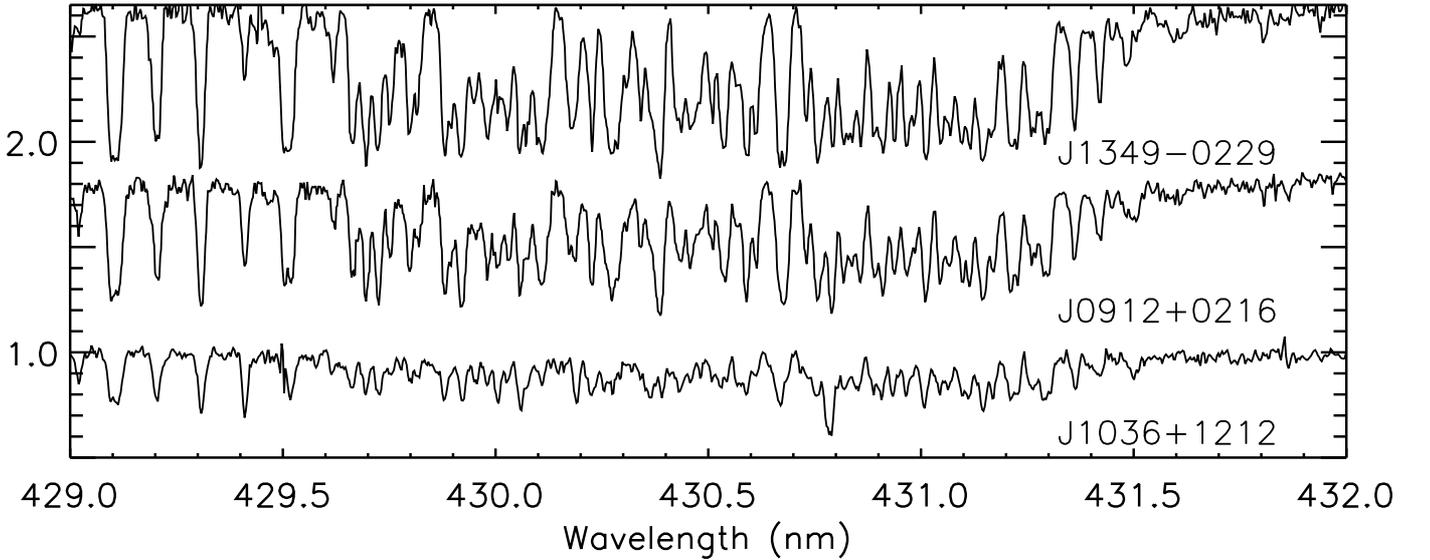}
\caption{Observed UVES spectra of the CH {\it G} band in the three CEMP
  stars. Spectra have been offset for clarity.}   
\label{fig1}
\end{figure*}

We observed 17 candidates; among these were the objects SDSS J1349-0229, SDSS J0912+0216
and SDSS J1036+1212. The SDSS spectra showed strong $G$ bands in these stars,
and the weakness of the Ca\,{\sc ii} K lines indicated very low
metallicities. Figure~\ref{fig1} displays the prominent CH {\it G}
band of all three stars. This band is hardly detectable in non-C-enhanced
stars of this effective temperature and metallicity.
We thus targeted the three objects as CEMP stars. An analysis of the remaining
  14 candidates will be presented in a future paper (Bonifacio et al.~2010).

The slit was set to 1\farcs{4} and the detector
was read with a $2\times2$ on-chip binning. 
The resulting  resolution is $R\sim 30\,000$. Supplementary observations were
obtained for SDSS J1349-0229 \relax
in the ESO period 83 to study the UV OH lines and the
permitted \ion{O}{i} triplet at 777 nm. Details on the settings and
exposure times are listed in Table~\ref{tab_obs} along with measurements of
the radial velocities. The second epoch observation of SDSS J1349-0229 reveals
a large radial-velocity variation of the order of 30 \kms, indicating clearly
that it is a member of a binary system. The signal-to-noise ratio
  varies from 35 at 392 nm up to 110 above 650 nm.

\section{Atmospheric parameters}

The effective temperatures of the stars were determined using both the wings of
H$\alpha$ and the Fe {\sc i} excitation equilibrium. 
The  surface gravity
was derived from the Fe {\sc i}/Fe  {\sc ii} ionisation equilibrium, while the
microturbulence was determined from Fe {\sc i} lines. The FITLINE code
(Fran\c{c}ois et al.~2003) was used to measure the equivalent widths of the Fe
{\sc i} and Fe {\sc ii} lines. ATLAS model atmospheres and SYNTHE (Kurucz 1993, 2005b)
synthetic spectra in their Linux version (Sbordone et al.~2004, Sbordone 2005) 
have been employed in the analysis, using scaled solar abundances with the
exception of the alpha-elements, which were enhanced by 0.4 dex. 
The opacity distribution functions of \citet{CK03} with 
a microturbulence of 1 \kms have been used.
The H$\alpha$ profiles were computed using the ATLAS9 
models and a modified
version of the BALMER code\footnote{Provided by R. Cayrel and
C.  Van't Veer, the original BALMER code of
R.L. Kurucz is available  at \url{http://kurucz.harvard.edu/}}. 
The self-broadening of H$\alpha$ was treated with 
the \citet{barklem00,barklem00b} theory,
and for the Stark broadening the computations of \citet{stehle}
were used.
Adopted stellar
parameters are listed in Table~\ref{tab1}. 

The carbon and nitrogen abundances are significantly enhanced in these
stars. In order to investigate the impact of the carbon
abundance on the atmospheric structure, 
models were computed for our stellar parameters using 
ATLAS12 \citep{kurucz96,K05,atlas12} with a carbon
enhancement of +3.0 dex, and a nitrogen enhancement of +2.0 dex. 
The average temperature difference in the two temperature structures
  is of the order of 20\,K. This small change in the temperature
  distribution translates into an error of approximately 0.02 dex in
  the abundance determination.
The analysis was therefore performed using the ATLAS9 models. 

\begin{table}
\caption{Adopted stellar parameters. }
\begin{tabular}[t]{lcccc}
Star       & $T_{\rm eff}$ & log $g$  &  $\xi$  & [Fe/H]  \\
           &  K            &  cgs     & \kms    &  dex    \\
\hline
SDSS J1349-0229 & 6200        & 4.0     & 1.5     & --3.0  \\
SDSS J0912+0216 & 6500        & 4.5     & 1.5     & --2.5  \\
SDSS J1036+1212 & 6000        & 4.0     & 1.4     & --3.2  \\
\end{tabular}
\label{tab1}
\end{table}

\section{Abundances}

The majority of the elemental abundances presented were determined by the
equivalent widths of unblended lines. A 1D LTE analysis was performed with the
WIDTH9 code \citep{KCD13,K05,castelli2005}.  
The van der Waals damping constants have been taken
from the Kurucz data base when available. When not
available SYNTHE uses the WIDTH approximation
(see Castelli 2005b and also Ryan 1998).
Abundances of strong lines, blended lines, or
lines that are affected by hyperfine structure were 
treated by line-profile fitting, using the
same code as in \citet{caffau05}. 
The mean LTE 1D abundance for each of the 35 elements
investigated in the three stars in our sample are presented in
Table~\ref{tab_fullA}.  
In each line of the table we provide an estimate of the error
for each star. For abundance indicators
with multiple lines this is simply the standard deviation 
of the abundances derived from individual lines.
For complex molecular bands like C$_2$ the error
has been estimated from the $\chi^2$ of the
best fitting synthetic spectrum. 
The systematic errors for changes in 
effective temperature and surface gravity are similar to those 
of \citet{sivarani}: 0.10--0.15\, dex 
in [X/Fe] for a change of 100\,K in effective
temperature or 0.5\,dex in log $g$.
As a reference we adopted the solar abundances
derived with the use of the CO$^5$BOLD 
solar model, when available \citep{CLS},
for other elements we adopted the values of
\citet{lodders}.

\begin{table*}
\caption{Elemental abundances for the three CEMP stars obtained from an
  analysis using 1D stellar atmospheres.}
\label{tab_fullA}  
\centering                          
\begin{tabular}{l| ccc | ccc | ccc | ccc}
\hline
                  & \multicolumn{3}{c}{SDSSJ1349-0229}   & \multicolumn{3}{c}{SDSSJ0912+0216}   & \multicolumn{3}{c}{SDSSJ1036+1212}   & Sun \\
\hline 
Element           & log($\epsilon$) & [X/Fe] & $\sigma$ & log($\epsilon$) & [X/Fe] & $\sigma$ & log($\epsilon$) & [X/Fe] & $\sigma$ & log($\epsilon$) \\
\hline
Li                &        &    &       &      &      &       & 2.21 &      &       &  \\
C(CH)             & 8.32 & 2.82 & 0.05  & 8.17 & 2.17 & 0.03  & 6.77 & 1.47 & 0.10  & 8.50 \\
C(C2)             & 8.66 & 3.16 & 0.02  &      &      &       &      &      &       & 8.50 \\
C(C {\sc i}) NLTE & 7.92 & 2.42 & 0.11  & 7.38 & 1.38 & 0.05  &      &      &       & 8.50 \\
N(NH)             & 6.46 & 1.60 & 0.12  & 7.11 & 1.75 & 0.12  & 5.95 & 1.29 & 0.25  & 7.86 \\
O(OH)             & 7.64 & 1.88 & 0.10  &      &      &       &      &      &       & 8.76 \\
O(O {\sc i}) NLTE & 7.39 & 1.63 & 0.14  &      &      &       &      &      &       & 8.76 \\
Na {\sc i} NLTE   & 4.79 & 1.49 & 0.01  & 4.18 & 0.38 & 0.03  & 3.53 & 0.43 & 0.01  & 6.30 \\
Mg {\sc i}        & 5.12 & 0.57 & 0.16  & 5.26 & 0.21 & 0.13  & 4.35 & 0.00 &  0.06 & 7.55 \\
Al {\sc i} NLTE   & 3.51 & 0.11 & ...   & 4.08 & 0.18 & ...   & 3.24 & 0.04 & ...   & 6.46 \\
Ca {\sc i}        & 3.74 & 0.40 & 0.08  & 3.32 & 0.42 & 0.12  & 3.52 & 0.38 & 0.17  & 6.34 \\
Sc {\sc ii}       & 0.08 & 0.01 & ...   & 0.85 & 0.28 & ...   &--0.02& 0.11 & 0.15  & 3.07 \\
Ti {\sc i}        &      &      &       & 2.96 & 0.54 & 0.03  &      &      &       & 4.92 \\
Ti {\sc ii}       & 2.47 & 0.55 & 0.09  & 2.93 & 0.51 & 0.18  & 2.47 & 0.75 & 0.26  & 4.92 \\
Cr {\sc i}        & 2.61 &--0.02& 0.21  & 2.99 &--0.16& 0.06  & 2.33 &--0.12& 0.17  & 5.65 \\
Cr {\sc ii}       & 2.82 & 0.17 & ...   & 3.07 &--0.08& 0.21  & 2.43 &--0.02& 0.30  & 5.65 \\
Mn {\sc i}        & 1.82 &--0.68& 0.08  & 2.45 &--0.55& ...   &      &      &       & 5.50 \\
Mn {\sc ii}       &      &      &       & 2.49 &--0.51& 0.03  & 1.84 &--0.46& 0.20  & 5.50 \\
Co {\sc i}        & 2.24 & 0.33 & 0.19  & 2.72 & 0.31 & 0.15  & 2.28 & 0.57 & 0.09  & 4.91 \\
Ni {\sc i}        & 3.48 & 0.26 & 0.37  & 3.79 & 0.07 & 0.18  & 3.30 & 0.28 & 0.22  & 6.22 \\
Sr {\sc ii}       & 1.21 & 1.30 & 0.17  & 0.98 & 0.57 & 0.07  &--0.85&--0.56& 0.04  & 2.91 \\
Y  {\sc ii}       & 0.49 & 1.29 & 0.29  & 0.31 & 0.61 & 0.20  &--0.76& 0.24 & 0.43  & 2.20 \\
Zr {\sc ii}       & 1.16 & 1.56 & 0.32  & 1.18 & 1.08 & 0.17  & 0.42 & 1.02 & 0.17  & 2.60 \\
Ru {\sc i}        &      &      &       & 1.93 & 2.61 & ...   &      &      &       & 1.82 \\
Pd {\sc i}        &      &      &       &      &      &       & 0.90 & 2.40 & 0.19  & 1.70 \\
Ba {\sc ii}       & 1.35 & 2.17 & 0.12  & 1.17 & 1.49 & 0.08  & 0.15 & 1.17 & 0.15  & 2.18 \\
La {\sc ii}       &--0.08& 1.74 & 0.05  & 0.03 & 1.35 & 0.14  & 0.37 & 2.39 & 0.60  & 1.18 \\
Ce {\sc ii}       & 1.24 & 2.63 & 0.11  & 1.28 & 2.17 & 0.14  & 0.73 & 2.32 & 0.41  & 1.61 \\
Pr {\sc ii}       & 0.65 & 2.87 & 0.07  & 0.53 & 2.25 & 0.13  & 0.03 & 2.45 & 0.11  & 0.78 \\
Nd {\sc ii}       & 0.37 & 1.91 & 0.12  & 0.08 & 1.12 & 0.31  & 0.35 & 2.08 & 0.36  & 1.46 \\
Sm {\sc ii}       & 0.30 & 2.35 & 0.01  & 1.05 & 2.60 & 0.08  & 0.67 & 2.92 & 0.15  & 0.95 \\
Eu {\sc ii}       &--0.86& 1.62 & 0.08  &--0.78& 1.20 & 0.05  &--1.42& 1.26 & 0.24  & 0.52 \\
Gd {\sc ii}       & 0.56 & 2.50 & 0.23  & 1.36 & 2.80 & 0.08  & 0.48 & 2.62 & 0.44  & 1.06 \\
Tb {\sc ii}       & 0.00 & 2.69 & 0.40  & 0.46 & 2.64 & 0.36  & 0.01 & 2.90 & 0.24  & 0.31 \\
Dy {\sc ii}       & 0.52 & 2.39 & 0.17  & 0.59 & 1.96 & 0.13  & 0.39 & 2.46 & 0.34  & 1.13 \\
Er {\sc ii}       & 0.68 & 2.73 & 0.08  & 0.48 & 2.03 & 0.23  & 0.61 & 2.86 & 0.04  & 0.95 \\
Tm {\sc ii}       &      &      &       &      &      &       &--0.31& 2.78 & 0.13  & 0.11 \\
Hf {\sc ii}       & 1.01 & 3.14 & ...   & 1.09 & 2.72 & 0.19  & 0.02 & 2.35 & 0.60  & 0.87 \\
Os {\sc ii}       &      &      &       &      &      &       & 0.28 & 2.11 & ...   & 1.37 \\
Ir {\sc i}        &      &      &       &      &      &       & 0.15 & 2.00 & ...   & 1.35 \\
Pb {\sc i}        & 2.14 & 3.09 & ...   & 1.88 & 2.33 & 0.30  &      &      &       & 2.05 \\
\hline
\end{tabular}
\end{table*}


It is known that convection in metal-poor stars induces very low temperatures
in the outermost layers, 
which are not predicted by classical 1D stellar atmospheres 
\citep{asp99,collet2007,CL07,jonay}, this behaviour
is often referred to as ``overcooling''. 
This is particularly important for molecular lines,
which form high up in the atmosphere, where the overcooling is 
largest. Investigations of CH, NH and OH  
abundances in dwarf atmospheres with [Fe/H] = --2.0 by Asplund (2004) have
shown that 1D models tend to overestimate abundances by more than 0.3 dex
when compared to 3D hydrodynamical simulations. Collet et al.~(2007) performed
a similar investigation with red giant stars at [Fe/H] = --3.0 and found the
corrections for CH, NH and OH to be in the range of --0.5 to --1.0 dex.

It is thus mandatory to use
3D hydrodynamical simulations
to assess molecular abundances in metal-poor stars.
For this purpose, we employed 3D model atmospheres computed with the 
CO$^5$BOLD code (Freytag
et al.~2002; Wedemeyer et al.~2004). 
The 3D spectral synthesis calculations
were performed with the code 
{\sf Linfor3D}\footnote{http://www.aip.de/$\sim$mst/Linfor3D/linfor\_3D\_manual.pdf}. 
The parameters of the 3D models used in the analysis are listed in
Table~\ref{tab_3dmod}. The models were 
chosen from our 3D model grid \citep{ludwigjd10} to best 
correspond with the
stellar parameters of our three stars. 
Since we did not have a 3D model with [Fe/H]~=~--2.5,
we interpolated for SDSS J0912+0216 
between the corrections obtained from the
models with [Fe/H]~=~--2.0 and --3.0 (d3t65g45mm20n01 \& d3t65g45mm30n01).

For each full 3D model,  {\sf Linfor3D} computes
a temporal and horizontal
average of the 3D structure over surfaces of equal 
(Rosseland) optical depth (hereafter 
denoted as $\left\langle\mathrm{3D}\right\rangle$). 
We compared the abundances derived
from each of our 3D models to this average model
$\left\langle\mathrm{3D}\right\rangle$ and to a corresponding standard 
hydrostatic 1D model atmosphere (1D$_{\rm LHD}$). The 1D$_{\rm LHD}$ models are
calculated assuming plane-parallel geometry and employ the same micro-physics
(equation-of-state, opacities) as CO$^5$BOLD. Convection is described by
a mixing-length theory in the formulation of
\citet{mihalas}; the adopted mixing-length-parameter
was 0.5. 

Two main effects distinguish 3D models from 1D models: the average
temperature profile and the horizontal temperature fluctuations.
The contribution of both effects is quantified by the 3D correction in the
sense 3D -- 1D. The correction attributed solely to the horizontal temperature
fluctuations is denoted by 3D -- $\left\langle\mathrm{3D}\right\rangle$. 
A complete description of 3D
corrections is given by Caffau et al.~(2008) and references therein.

\subsection{Lithium}

The Li {\sc i} 670.7 nm doublet was detected in SDSS J1036+1212, and an abundance of A(Li) =
2.21 was derived. The doublet is not detected in SDSS J1349-0229 and SDSS
J0912+0216. 

\begin{table}
\caption{Parameters of 3D models used in the analysis.} 
 \begin{tabular}[t]{lrlrr}
Model       & $T_{\rm eff}$ & log $g$  &  [Fe/H] \\
\hline
d3t63g40mm30n01 & 6270    & 4.00     &  --3.0  \\
d3t65g45mm20n01 & 6530    & 4.50     &  --2.0  \\
d3t65g45mm30n01 & 6550    & 4.50     &  --3.0  \\
d3t59g40mm30n02 & 5850    & 4.00     &  --3.0  \\
\end{tabular}
\label{tab_3dmod}
\end{table}

\begin{figure}
   \includegraphics[width=0.5\textwidth]{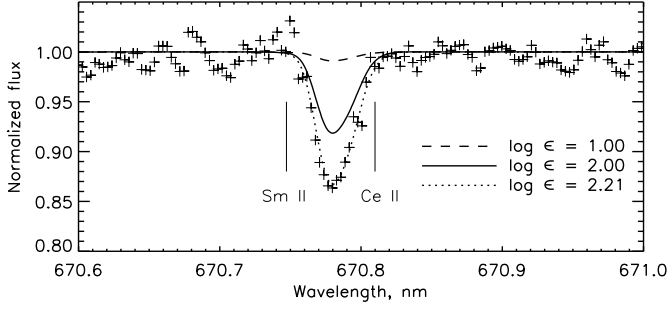}
   \caption{Observed and synthetic spectra of SDSS J1036+1212 in the region of
   the Li {\sc i} 670.7 nm doublet. Nearby lines of Ce {\sc ii} and Sm {\sc
     ii} were included in the synthesis. Synthetic spectra are shown with
   lithium abundances of $\epsilon$ = 1.00, 2.00 and 2.21. See the text for
   details.  \label{Li}} 
  \end{figure}

The abundance of Li determined is quite high for this type of star, but
``normal'' for dwarf stars of this metallicity. In 2002,
Reyniers et al. presented an alternative identification of the Li doublet in 
post-AGB stars: a Ce {\sc ii} line. Since SDSS J1036+1212 is strongly enhanced
in neutron-capture elements 
(see Sect. 4.8), we investigated this possibility.
The observed line is plotted in Fig.~\ref{Li}. Synthetic spectra were
calculated with the abundance of Sm {\sc ii} and Ce {\sc ii} listed in
Table~\ref{tab_fullA} and three abundances of lithium: log $\epsilon$ = 1.00, 2.00
and 2.21. 

Even with a significant enhancement, the contribution of the Sm 
{\sc ii} and Ce {\sc ii} lines are minimal. A stronger contribution from these
lines would require a much lower surface gravity at this temperature.
Additionally, we do not find that the line is shifted redwards compared to
nearby lines, as was the case for the AGB stars of Reyniers et al.~(2002).
We can therefore confirm the identification
of the observed feature with the \ion{Li}{i} 
resonance doublet and a lithium abundance of 2.21 for this star at our
stellar parameters.  

\subsection{Carbon}

Although in principle the carbon abundance of CEMP stars may be determined
from both molecular and atomic lines, the majority of the values cited in
the literature are derived from spectral lines of CH in the $G$ band, as these
are strong features in CEMP star spectra. In some cases C$_2$ features and
C{\sc i} lines may be present. We detect CH, C$_2$ and C{\sc i} lines in SDSS
J1349-0229 and CH and C{\sc i} lines in SDSS J0912+0216. These stars exhibit a very
strong carbon enhancement. In SDSS J1036+1212, the carbon enhancement is much
weaker, and only CH lines could be used to determine [C/Fe]. The results are
summarised in Table~\ref{tab_cn}. 

In order to perform 3D spectrum synthesis
of molecular lines, which are computationally
demanding,  we decided to focus on four
isolated CH features at 416.4 nm, 416.9 nm, 418.0
nm, and 418.8 nm  to determine [C/Fe] in all three stars.
The line data  were taken from the molecular
line lists of Kurucz \citep{Kweb},  but the 
oscillator strengths were scaled by a 
factor of 0.4, as was done by \citet{bonifacio1998} to place
the log $gf$ values on the same scale as the ones by \citet{norris1997}.
From our 1D analysis we obtain [C/Fe] = 2.82 $\pm$ 0.05, 2.17 $\pm$ 0.03 and 1.47 $\pm$ 0.10 for SDSS J1349-0229, 
SDSS J0912+0216 and SDSS J1036+1212, respectively. The 3D corrections for
these lines are quite substantial, have the effect to reduce the carbon
abundance, and range from --0.73 to --0.50 dex.  

\begin{figure}
   \includegraphics[width=0.5\textwidth]{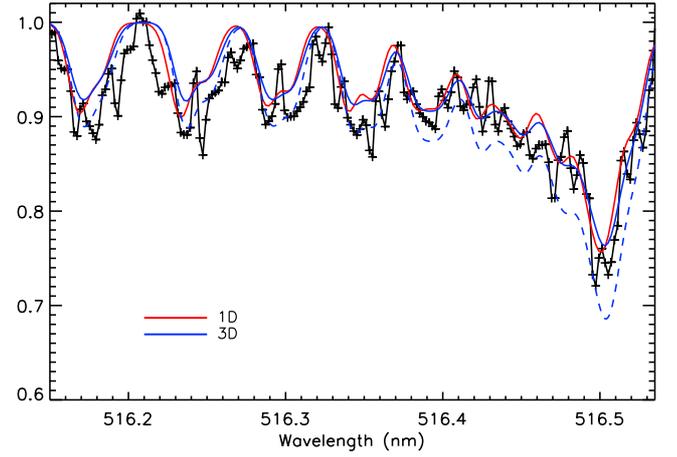}
   \caption{
   \label{c2lines} 
Observed and 1D and 3D synthetic spectra of SDSS J1349-0229 in the region of
   the C$_2$ band at 516.3 nm. The 3D spectra was computed with
   [C/Fe]=1.72 while the 1D spectra was computed with [C/Fe]=3.16.
   The dashed line shows the 3D synthetic spectra
   with an increase of 0.15 dex in carbon abundance.}
  \end{figure}

A 1D carbon abundance of [C/Fe] = 3.16 was derived from C$_2$ features at
516.3 nm in the spectrum of SDSS J1349-0229. 
As for CH, we used the molecular line lists of Kurucz  \citep{Kweb} for the
C$_2$ lines.  
The 3D correction was computed from a synthesis of 118 lines centered at 516.3
nm and spanning 0.6 nm. Observed and synthetic spectra of the feature are
shown in Fig.~\ref{c2lines}. The 3D corrections for this feature are much
more substantial than the corrections for the CH lines,
they even reach --1.44 dex. This is because the C$_2$ lines form in layers which are
more superficial than those where the CH lines 
form, where the effects of overcooling and photospheric
  inhomogeneities are strongest.

Three C{\sc i} lines were measured in SDSS J1349-0229 and two in SDSS J0912+0216. All of
these lines are high excitation lines, and are highly sensitive to NLTE
effects. Using  LTE  largely 
overestimates the abundance. The 3D corrections on the other hand, show that the 1D abundance
is underestimated, however by a very small amount, the largest corrections
being 0.08 dex. NLTE corrections were calculated using the Kiel code
\citep{SH} and the carbon 
model atom described in  \citet{stuerenburg}, 
and are listed in Table~\ref{tab_cNLTE}.
The effects of collisions with
hydrogen atoms were treated as described
in \citet{SH}, who generalised the \citet{Drawin}
formalism.
We provide NLTE corrections 
 for three different
values of S$_H$, a scaling factor, S$_H=0$ 
means totally neglecting collisions with H atoms,
while S$_H=1$ means
taking the value of the \citet{SH} formalism. 
The NLTE C{\sc i} abundance listed in  
Table~\ref{tab_abund} adopts the NLTE corrections for S$_H$=1/3.

The three abundance indicators for SDSS J1349-0229 do not give consistent
results either
in the 1D or in the 3D analysis. 
The standard deviations of the abundances from the different
  features are 0.30 and 0.32 from the 1D and the 3D analyses, respectively.
In 1D the most significant difference occurs between the C$_2$ and
C{\sc i} lines, where we find a difference of 
0.74 dex. In 3D the biggest difference of 0.79\,dex 
occurs between the C{\sc i} and
C$_2$ lines. The C{\sc i} lines give a much higher abundance. Meanwhile the
difference between 3D CH and C{\sc i} is 0.42 dex, assuming an S$_H$=1/3.  
For SDSS J0912+0216, where only CH and C{\sc i} lines are available, we
find differences of 0.79 and 0.23\,dex and $\sigma$ = 0.40 and 0.12 in 1D and
  3D, respectively.

\begin{figure}
   \centering
   \includegraphics[width=0.5\textwidth]{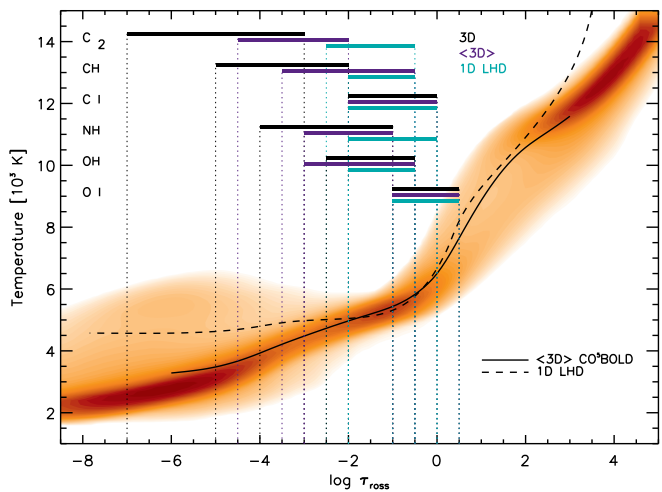}
   \caption{
The temperature structure for the 3D model d3t63g40mm30n01 is
     plotted as a function of $\tau_{\rm ross}$ along with the average 3D ($\left\langle\mathrm{3D}\right\rangle$) and the
     corresponding 1D$_{\rm lHD}$ temperature structures. Overplotted on the figure
     are the ranges of the depth of formation of the C$_2$, CH, C{\sc i}, NH,
     OH and O{\sc i} spectral lines used in the analysis.}
   \label{temp_struct} 
  \end{figure}

The sensitivity of the molecular lines to temperature is clear from the
significant 3D corrections. In
Fig.~\ref{temp_struct} we have plotted the temperature distributions for the
3D, $\left\langle\mathrm{3D}\right\rangle$, and 1D$_{\rm LHD}$ models. The
ranges of the depth of formation, intended
as the range in optical depth over which the line
contribution function is significantly different from zero, 
of the CH, C$_2$ and C{\sc i} lines are
overplotted as horizontal lines. 
In 1D we derive a larger C abundance
from the C$_2$ lines than from the CH lines, while in 3D the
reverse is true. This can be understood by looking
at the
different depths of formation of the 
C$_2$ and CH lines in the 1D and 3D models.
The C$_2$ lines are formed higher up in the
atmosphere compared to the CH lines, but 
far more in the 3D model than in the 1D. 
The 3D models do not achieve a better consistency between
CH and C$_2$ lines.
The C {\sc i}
lines are quite insensitive to 3D effects. 
Neither in 3D nor in 1D
we achieved consistent results between molecular lines
and C{\sc i}. 

In view of the high formation region of
C$_2$ we must consider the validity of the 
models in this region of the atmosphere. 
The 3D model properties, like the choice of the opacity binning scheme, may
well have an effect on the structure of the outer regions of the atmosphere. To
explore this effect, we recomputed the carbon 3D corrections for 
SDSS J1349-0229 with a 3D model atmosphere that employs 12 opacity bins versus the
six bins used in the models listed in Table~\ref{tab_3dmod}. The 3D corrections
are compared in Table~\ref{tab12bins}. The 12 bin model gives a better agreement between the CH 
the C$_2$ lines, and both molecular features
appear agree better with C {\sc i}. 
Although these results appear encouraging, 
the formation region of the C$_2$ lines is quite high and at low density.
The validity of LTE for
molecule formation and levels population may well be an oversimplification.
For further discussion on this topic see Behara et al.~(2009).

  \begin{figure}
   \centering
   \includegraphics[width=0.5\textwidth]{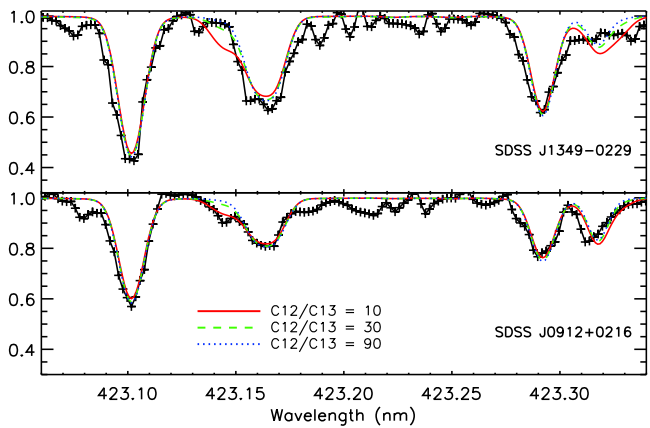}
   \caption{Observed and synthetic spectra for two stars in the region of
     the $^{13}$CH and $^{12}$CH lines. Solid, dashed and dotted lines show
     the synthetic spectrum with a ratio of $^{12}$C/$^{13}$C = 10, 30
     and 90 respectively.}
   \label{fig_13CH} 
  \end{figure}

\subsubsection{The $^{12}$C/$^{13}$C ratio}

CH features from the $R$-branch of 
the $A^{2}\Delta - X^{2}\Pi$ band
were used in our analysis of the
$^{12}$C/$^{13}$C isotopic ratio.
Only lower limits could be derived:
30 for SDSS J1349-0229, and 10 for SDSS J0912+0216.
Observed spectra and fits to synthetic spectra are shown in
Fig.~\ref{fig_13CH}. The low carbon abundance of SDSS
J1036+1212 prevented us from obtaining a lower limit to 
the $^{12}$C/$^{13}$C isotopic ratio.

\subsection{Nitrogen}

We obtain the nitrogen abundances for the three stars from isolated NH
features at $\sim$ 336 nm. We used the Kurucz data 
\citep{Kweb} for the NH molecule in our
analysis, and following Spite et al.~(2005) applied a correction of --0.40
dex to our abundances. 
The 1D analysis yields [N/Fe] = 1.60, 1.75 and
1.29 for SDSS J1349-0229, SDSS J0912+0216 and SDSS J1036+1212 respectively. 
The magnitude of the 3D corrections for NH is greater than the corrections
for CH in all three stars and follows the same direction by reducing the 1D
abundance. The corrections range from --0.67 to --0.93. The adopted 3D
abundances are listed in Table~\ref{tab_cn}.

\subsection{Oxygen}

The oxygen forbidden lines at 630.0 and 636.4 nm are not detected in the
spectra of our stars. Measured upper limits range from [O/Fe] =
  2.0 to 2.5 for the three stars. Spectra in the range of the UV OH lines and the
permitted \ion{O}{i} triplet at 777 nm were only available for SDSS
J1349-0229, and therefore we can only determine the oxygen abundance for this
star. We computed the abundance from three isolated UV  
OH lines of the (0-0)
vibrational band of the $A^2\Sigma - X^2\Pi$ electronic systems, and computed
the corresponding 3D corrections. The $gf$ values of the OH lines 
were computed from the lifetimes calculated by Goldman \& Gillis (1981).
The result is [O/Fe] = 1.70. We obtained a
higher abundance from the triplet lines: [O/Fe] = 1.81. However, it is well
known that these high excitation lines suffer from NLTE effects 
(see e.g. Gratton et al.~1999 and references therein). 
We  computed NLTE corrections using the Kiel
code \citep{SH} and the model-atom described in
\citet{paunzen}. For oxygen the collisions
with hydrogen atoms were treated with the
\citet{SH} formalism for three choices of
S$_H$. The corrections are listed in
Table~\ref{tab_cNLTE}. Taking this into account, we obtained an abundance from
the triplet lines of 1.69, which agrees excellently with the abundance obtained from
the OH lines.   

\begin{figure}                                 
   \centering
   \includegraphics[width=0.50\textwidth]{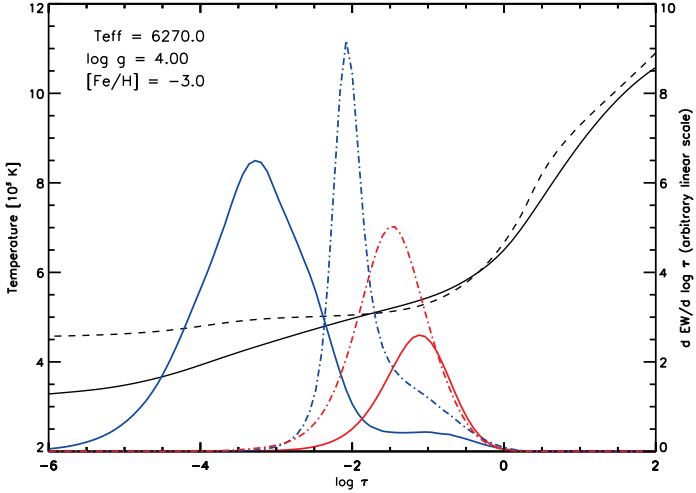}
      \caption{
  Equivalent width contribution function plotted as a
  function of optical depth for the 3D model (blue) and the 1D$_{\rm LHD}$
  model (red) for two different C/O ratios. Scaled solar
  carbon-to-oxygen is plotted as a solid line, while [C/O] = 1 is plotted as a
  dot-dashed line. Overplotted are the temporal and horizontal
  average of the temperature profile of the 3D model (black solid line) and the temperature
  profile of the 1D$_{\rm LHD}$ model (dashed black line).
}
         \label{Fig_3DOH}
   \end{figure}

The 3D corrections for the OH lines are quite low ($\sim -0.2$)
compared to values found in the literature for metal-poor stars: Asplund \&
Garc\'ia P\'erez (2001) state values of less than --0.60 dex, and \citet{jonay}
obtain corrections of --1.50 dex for dwarfs stars with [Fe/H] = --3.00. 
The 3D corrections for OH lines are very sensitive to 
the C/O ratio in the atmosphere. We fixed [C/O] = 1.00 \relax in our
calculations. Using a solar carbon-to-oxygen ratio yields a correction of
--1.35 dex. Figure~\ref{Fig_3DOH} shows the contribution
function for a line computed for the two different C/O ratios. When carbon is
enhanced, there is no contribution to the OH line higher in the atmosphere,
since the oxygen is tied up in CO in this region due to the high carbon content. 

\begin{table}
\caption{Carbon, nitrogen and oxygen abundances and 3D corrections. Abundances
  are expressed with respect to the solar abundances of log\,($\epsilon$)\,=\,8.50,
  7.86, and 8.76 for carbon, nitrogen and oxygen, respectively. }
\label{tab_abund}  
\begin{tabular}{l cccc}
\hline 
Element      & [X/Fe]$_{\rm 1D}$ & [X/Fe]$_{\rm 3D}$ & 3D--1D  &  3D--$\left\langle\mathrm{3D}\right\rangle$ \\
\hline
             & \multicolumn{4}{c}{SDSS J1349-0229}   \\
CH           & 2.82 & 2.09  & --0.73 & --0.37 \\
C$_2$        & 3.16 & 1.72  & --1.44 & --0.82 \\
C {\sc i} NLTE   & 2.42 & 2.51 & 0.08   & --0.04 \\
NH           & 1.60 & 0.67  & --0.93 & --0.34 \\
OH           & 1.88 & 1.70  & --0.18 & --0.06 \\
O {\sc i} NLTE   & 1.63 & 1.69  & 0.06   & --0.04 \\
\hline
             & \multicolumn{4}{c}{SDSS J0912+0216}  \\
CH           & 2.17 & 1.67  & --0.50 & --0.20 \\
C {\sc i} NLTE   & 1.38 &  1.44 &  0.06  & --0.03 \\
NH           & 1.75 & 1.07  & --0.67 & --0.30 \\
\hline
             &  \multicolumn{4}{c}{SDSS J1036+1212}   \\
CH           & 1.47 & 0.96  & --0.51 & --0.11 \\
NH           &  1.29 & 0.51  & --0.78 & --0.08 \\
\hline
\end{tabular}
\label{tab_cn}
\end{table}

\begin{table}
\caption{NLTE corrections for carbon and oxygen lines.}
\begin{tabular}{l ccc }   
\hline
Line (nm)           & S$_H$ = 1.0   & S$_H$ = 1/3   & S$_H$ = 0.0   \\
\hline 
                    & \multicolumn{3}{c}{SDSS J1349-0229}   \\
C {\sc i} 493.2     & --0.352    & --0.432    & --0.502 \\
C {\sc i} 505.2     & --0.369    & --0.456    & --0.535 \\
C {\sc i} 538.0     & --0.379    & --0.468    & --0.543 \\
O {\sc i} 777.1     & --0.127    & --0.144    & --0.130 \\ 
O {\sc i} 777.4     & --0.098    & --0.111    & --0.099 \\
O {\sc i} 777.5     & --0.100    & --0.113    & --0.101 \\
\hline
                    & \multicolumn{3}{c}{SDSS J0912+0216}   \\
C {\sc i} 505.2     & --0.267    & --0.352    & --0.456 \\
C {\sc i} 658.8     & --0.258    & --0.335    & --0.425 \\
\hline
\end{tabular}
\label{tab_cNLTE}
\end{table}

 \begin{table}
 \caption{3D corrections calculated using 6 and 12 opacity bins.}
 \label{tab12bins}
 \begin{center}
 \begin{tabular}{r|ccc}
  3D model                 &  CH & C$_2$ & C {\sc i} {\small NLTE} \\
 \hline
 6 bin [X/Fe]$_{\rm 3D}$    & 2.09 & 1.72 & 2.51 \\
 12 bin [X/Fe]$_{\rm 3D}$   & 2.22 & 1.99 & 2.50 \\
 \end{tabular}
 \end{center}
 \end{table}

\subsection{The alpha elements (Z $>$ 8)}

Our measurements of [Mg/Fe], [Ca/Fe] and [Ti/Fe] in all three stars yield
results that are consistent with the majority of metal-poor stars 
\citep{cayrel04,B09}. 

\subsection{The odd-Z elements}

Our measurement of [Na/Fe] was made using the Na D resonance lines at 588.995
and 589.592 nm. We have applied NLTE corrections from Gratton et
al.~(1999). The corrections range from --0.07 to --0.15 dex.
All stars show an enhancement of Na, which is
typical for CEMP stars \citep{Stancliffe}.

The abundance of Al was measured from the 396.152 nm line.
We found [Al/Fe] to be nearly solar in all three stars after applying the NLTE
correction from Baum\"uller \& Gehren (1997) of +0.64, the mean value  given
for stars with [Fe/H] $<$ --2.20 in their Table 2. This
  correction agrees closely with the NLTE correction given by
  Andrievsky et al.~(2008) of 0.6 dex for stars with [Fe/H] = --3.00.

\subsection{The iron-peak elements}

The values of [Sc/Fe], [Mn/Fe] and [Co/Fe] all agree with typical values for
metal-poor stars \citep{cayrel04,B09}. The abundance of Mn and Co were
calculated using spectrum synthesis to take the effects of hyperfine splitting
(HFS) into account. In the analysis we adopt the hfs line lists
  of \citet{Kweb}.

We were able to measure Cr {\sc i} and Cr {\sc ii} lines in all stars. The
abundances derived from the Cr {\sc ii} lines were found to be systematically
higher than the values obtained from the Cr {\sc i} lines. 
\citet{B09} already noted this behaviour and suggested it may
be due to NLTE effects.
On average, we found a difference of 0.12 dex. 
The results of Cr {\sc i} are consistent with the wide
range of reported Cr abundances from CEMP stars though.

We note that 
\citet{bergemann} and \citet{bergemann09}
suggest that Mn, Co and Cr are affected by significant
NLTE effects although no computations exist yet
at a metallicity of --3. In any case these corrections
will be similar for CEMP stars and ``normal'' stars.

A wide range of [Ni/Fe] values are found in CEMP stars. All our stars show an
enhancement of Ni, which ranges from +0.07 to +0.28 dex. 

\subsection{Neutron-capture elements}

The spectra of all three stars display a wealth of neutron-capture element
lines. All in all 21 elements are detected. Pb is detected in two stars, while in
the third, SDSS J1036+1212, third peak r-process elements are detected. 

\begin{figure}[ht]
\centering \includegraphics[width=0.45\textwidth]{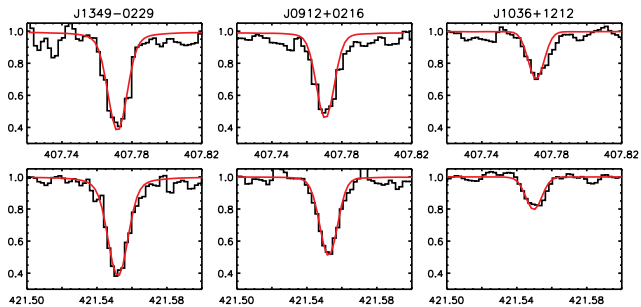}
\caption{Sr {\sc ii} 407.7 nm and 421.5 nm lines used in the
  abundance analysis.}
\label{Sr}
\end{figure}

We derive strontium abundances for our stars employing the strong Sr {\sc ii} 407.7 nm and
421.5 nm lines shown in Fig.~\ref{Sr}. SDSS J1349-0229 and SDSS J0912+0216 exhibit
overabundances of Sr of 1.30 and 0.57~dex, while SDSS J1036+1212 shows an
underabundance of Sr by 0.56 dex. Zirconium and yttrium are found to be
enriched in all stars. 

Barium is overabundant in all three stars, although slightly less so in SDSS
J1036+1212. The Ba abundances were estimated from the Ba {\sc ii} 493.4 nm and
614.1 nm lines. For the analysis we adopted the HFS by
McWilliam (1998) and assumed a solar isotopic mix. 
We note however that if a pure $r$-process isotopic mix is
  assumed, the barium abundance is decreased by 0.1 to 0.35 dex.

Abundances from lanthanum lines were derived in all stars with the hyperfine
structure constants and transition probabilities from Lawler et al.~(2001a).
The HFS components were calculated by the code LINESTRUC (Wahlgren 2005).

For the cerium abundance analysis we adopted the line list from Lawler et
al.~(2009). The praseodymium abundance was derived using line
lists from Li et al.~(2007) and Ivarsson et al.~(2001). Since the lines are
relatively weak, we ignored any HFS. The linelist from Den
Hartog et al.~(2003) was used to determine the neodymium abundance.

For the analysis of the samarium lines we applied the values from Lawler et
al.~(2006). Europium lines were detected in all stars, and abundances 
were determined with the help of spectrum synthesis and line profile fitting, 
taking into account the HFS. 
Gadolinium line data were taken from Den Hartog et al.~(2006), and data
for terbium from Lawler et al.~(2001b). 

We ignored HFS in dypsrosium as the lines were weak. Oscillator
strengths were adopted
from Wickliffe et al.~(2000). Data for erbium lines were taken
from Lawler et al.~(2008). For thulium we used the data from Wickliffe \&
Lawler (1997). The effects due to HFS are small and were
ignored. For the analysis of hafnium, we adopted values from Lawler et
al. (2007). 

Osmium and iridium lines were detected only in SDSS J1036+1212. 
Lead was detected in SDSS J1349-0229 and SDSS
J0912+0216. The line list from Aoki et al.~(2002) was used in the
analysis. 

\section{Comparison with similar stars}

It is now widely accepted that the CEMP class includes
objects which have rather diverse astrophysical origin.
In Sect. \ref{intro} we described
the classification scheme proposed by
Beers \& Christlieb~(2005) based on the
abundances of neutron-capture elements in 
CEMP stars. 
This classification is non-quantitative in the
sense that one refers loosely to 
``enhancement'' of $r$ or $s$-process elements 
without specifying which ones, and without
providing quantitative limits for
the enhancements.
In the note to their Table 4, \citet{sivarani}
attempted to provide a more quantitative description of the
classes, based essentially on the abundance of
Ba and for one class (CEMP-s) also on the Ba/Eu ratio.

Jonsell et al.~(2006) defined the class CEMP-r+s, which is different from the
class CEMP-r/s of Beers \& Christlieb (2005). 
In Table \ref{cemp_class} we provide the definitions found
in the literature of the different classes.
All stars are assumed to have [C/Fe]$>+1$.

\begin{table}[h]
\caption{Classification of CEMP stars based on abundances of neutron-capture
  elements. S2006 and J2006 refer to Sivarani et al.~(2006) and Jonsell et
  al.~(2006), respectively. }
\label{cemp_class} 
\begin{tabular}{lll}
Class       & Definition & Reference  \\
\hline
CEMP-no &  [Ba/Fe]$< 0.0$       & S2006  \\
CEMP-no/s &  $+0.5<$[Ba/Fe]$< +1.0$       & S2006 \\
CEMP-s &  [Ba/Fe]$> +1.0$ \& [Ba/Eu]$>+0.5$       & S2006 \\
CEMP-s &  [Ba/Fe]$> +1.0$ \& [Ba/Eu]$>+0.0$ & J2006 \\
       &  \& [Eu/Fe] $<+1.0$                & \\  
CEMP-r/s &  $0.0<$[Ba/Fe]$< +0.5$        & S2006 \\
CEMP-r+s &    [Ba/Fe] $>$ 1.0 \& [Eu/Fe] $>$ 1.0      & J2006 \\
\hline
\end{tabular}
\end{table}

The classification scheme above is phenomenological
and also leaves space for some ambiguity.
For instance, all three of our programme stars
belong to the CEMP-r+s class,
but  SDSS J1349-0229 also classifies as a CEMP-s (S2006) star.  
Moreover the really  striking characteristic
of the CEMP-no/s stars is that [Ba/Fe]$>0$ and
[Sr/Fe]$<0$; if we redefine the class
in this way, then  SDSS J1036+1212
would classify as a CEMP-no/s star.
Although our preferred carbon, nitrogen and oxygen abundances are
  those derived from the 3D analysis, we adopt the abundances from the
  1D analysis to compare our stars with a sample of hot dwarf
  CEMP stars from the literature.
The full sample of stars selected is compiled from Table 10 of Aoki et
al.~(2008), Table 8 of Jonsell et al.~(2006) and Table 4 of
\cite{sivarani}.

\subsection{Lithium}

Few values of lithium have been measured in CEMP stars. Often only upper
limits can be measured. 
Due to its fragility this element is an excellent
diagnostic of the thermal history of the material.
It is easily destroyed at temperatures above
$2.5\times 10^6$\,K, therefore material 
which has experienced these temperatures
should be essentially Li-free.
This could explain why CEMP stars
generally show no Li, as the temperatures
necessary to produce the carbon will
lead to Li destruction.

\begin{figure}[ht]
\centering \includegraphics[width=0.5\textwidth]{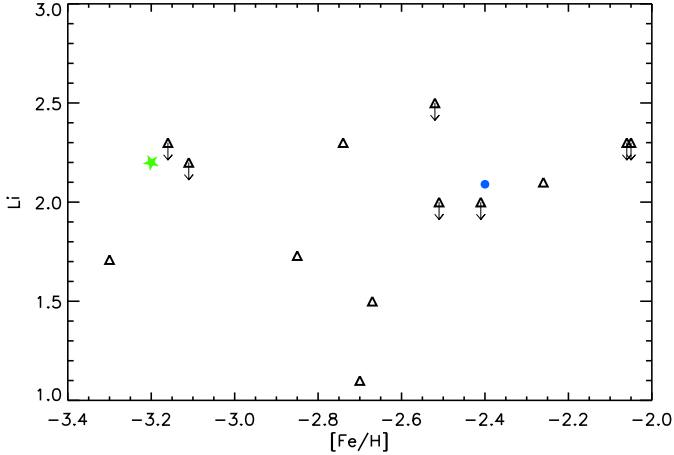}
\caption{Abundance of lithium as a function of [Fe/H]. Abundances from
  literature are plotted as open triangles, while the star symbol is from
  this work. The filled circle shows CS 22964-161.}
\label{li_fe}
\end{figure}

Still, a few CEMP stars have been observed to have Li values close to the
value of the Spite plateau \citep{spite82Natur,spite82}. 
We present a sample of measured abundances in
Fig.~\ref{li_fe}. 
A few stars show Li at the Spite plateau level, while
others show a lower Li abundance.
A remarkable example is the double lined spectroscopic
binary CS 22964-161 \citep{thompson2008}. The two stars are both CEMP, the
primary is a warm sub-giant, while the secondary is
a main sequence star. Li has been measured at the level
of the Spite plateau in the primary star and is 
only tentatively detected in the secondary, again close to
the Spite plateau.
If binary mass transfer is responsible for
the abundances of this system, it must once have been a triple system.
That Li is detected at all in a star polluted by nuclearly
processed material is in itself surprising, that it should
be at the level of the Spite plateau is even more.
There could of course be a conspiracy in a way that the Li produced in 
the donor star exactly  matches the destroyed Li, modulo the dilution
factor of the accreted material.
In the case of  CS 22964-161 one further requires
that the dilution factor has been very nearly the
same for both components of the binary.
Lithium may be produced via several mechanisms during the AGB phase. However,
if the observed lithium originates from the donor star, the question arises
why Li abundances {\em above} the Spite plateau have not been detected. 

We observe a Li abundance corresponding to the Spite plateau in SDSS
J1036+1212. 
A scenario alternative to mass transfer in a binary
star to explain the observed abundances
would be the formation of SDSS J1036+1212 from a cloud of C-rich material.     
Also in this case the C-rich material should be essentially 
Li-free though, thus the Li-puzzle remains.
Radial velocity monitoring of this star would be highly desirable to
constrain the existence of a binary companion.

\subsection{Carbon and nitrogen}

\begin{figure}[ht]
\centering \includegraphics[width=0.5\textwidth]{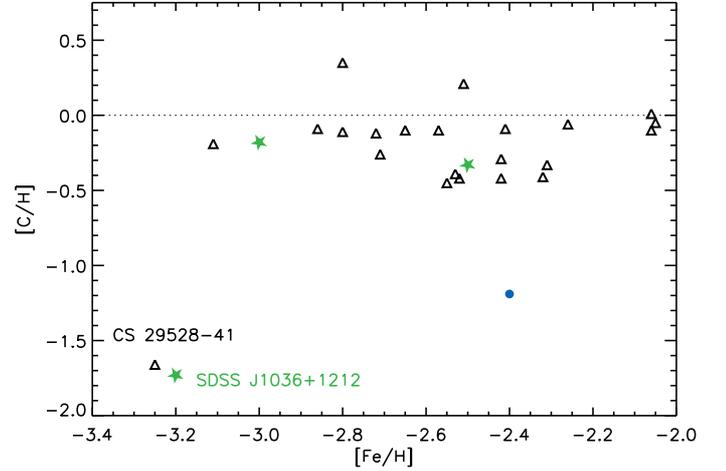}
\caption{Abundance of carbon as a function of [Fe/H]. Abundances from
  literature are plotted as open triangles, while star symbols are from
  this work. The filled circle shows CS 22964-161.}
\label{cefe}
\end{figure}

We plot [C/H] as a function of [Fe/H] in Fig.~\ref{cefe}. 
The open symbols are from the literature,
while the star symbols are from this work. Excluding three objects, SDSS
J1036+1212 (this work), CS 29528-041 (Sivarani et al.~2006) 
and CS 22964-161 \citep{thompson2008}, 
the stars display a rather constant [C/H], slightly less than solar. 
A possible explanation is that such  a roughly constant 
value is found when the accretion from the AGB star
occurs through Roche lobe overflow.
This suggests that the AGB companion
always provides basically the same mass of carbon
and the accretion mode through Roche lobe overflow 
ensures a roughly constant dilution factor.
In this scenario stars with low carbon abundance, like  SDSS
J1036+1212, CS 29528-041 and CS 22964-161 would arise 
through AGB wind instead from accretion, which implies larger
dilution fractions.

There are three problems with this scenario: the first is that one would 
expect the plane [C/H], [Fe/H] to be almost uniformly populated
with different efficiencies due to a mixture of cases of Roche lobe overflow accretion   
and wind accretion, but the few
``low'' carbon stars cluster around [C/H]$\sim -1.5$.
In the second place it is not at all clear that the carbon
yield of AGB stars should be constant.
Stars of different masses could provide different carbon abundances.
From the theoretical point of view,
while there are some discrepancies in the carbon yields for stars
of a mass lower than 3\,M$_\odot$, there seems to be
a general consensus that for higher mass AGB stars
the carbon yield is roughly constant  \citep{marigo,ventura,herwig2004,kl2007}.
Thus our scenario is acceptable only if CEMP stars 
arise predominantly in binary systems in which 
the primary is more massive than  3\,M$_\odot$.
It should be noted however that recently \citet{VM09}
pointed out that AGB yields need to be computed including the changes in surface yields and their 
effect on opacities. Such a treatment may lead to 
a substantial revision of published yields.
A third problem of our proposed scenario is that due to 
dilution; one should expect  ``low'' C stars to show
also lower  abundances of n-capture elements, due to the larger dilution,  
but this is not so compelling for the three 
stars shown in Fig.~\ref{cefe}. 
A larger sample of dwarf CEMP stars
at [Fe/H] $\leq$ --3.2 would be beneficial for a more complete discussion.

\begin{figure}
\centering \includegraphics[width=0.5\textwidth]{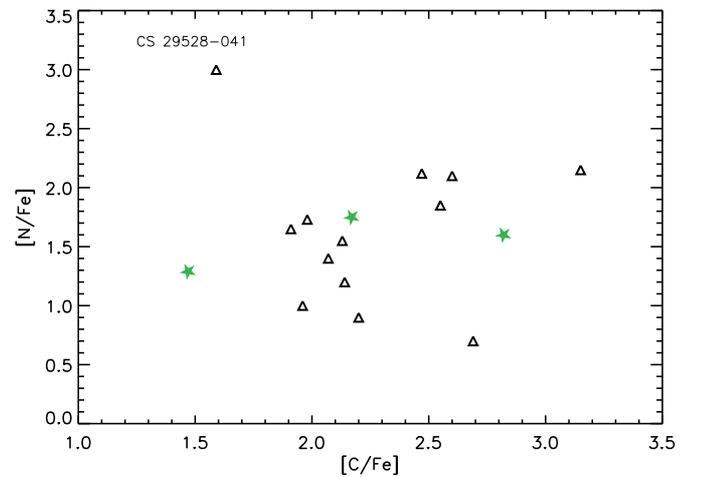}
\caption{Abundance of nitrogen as a function of [C/Fe]. Abundances from
  literature are plotted as open triangles, while star symbols are from
  this work.}
\label{nfe_cfe}
\end{figure}

The nitrogen abundances are shown in Fig.~\ref{nfe_cfe} as a function of
[C/Fe]. While CS 29528-041 exhibits a high N abundance for its low C
abundance, SDSS J1036+1212 displays a nitrogen abundance in the expected
range, as do the other two stars in this work. 

Our nitrogen abundances and lower limits on carbon isotopic
ratios are compatible with the predictions of
models of hot bottom burning in the envelopes of AGB stars 
(Karakas \& Lattanzio, 2007).
According to these authors
intermediate-mass low-metallicity AGB stars  produce low C/N and
$^{12}$C/$^{13}$C isotopic ratios, while low-mass AGB stars 
produce high C/N and $^{12}$C/$^{13}$C isotopic ratios. 
We measured [C/N] = 1.22 for SDSS J1349-0229 
and have a lower limit of 30 on its carbon isotopic
ratio, qualitatively compatible with a low-mass
AGB donor.
We found a much lower value, [C/N] = 0.42, for SDSS
J0912+0216, and  a lower limit on the  carbon isotopic
ratio of 10, compatible with an intermediate-mass 
AGB donor.

\subsection{Neutron-capture elements}

\begin{figure}[ht]
\centering \includegraphics[width=0.5\textwidth]{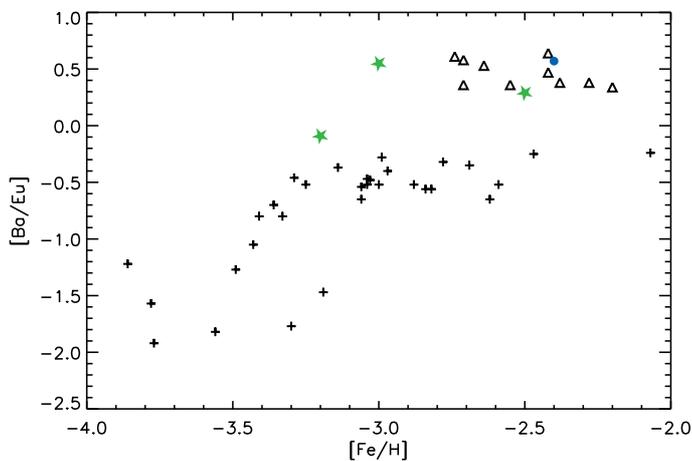}
\caption{[Ba/Fe] as a function of [Eu/Fe]. Abundances from
  literature are plotted as open triangles, while star symbols are from this
  work. Included in this plot are EMP stars from Fran{\c c}ois et al.~(2007),
  plotted as crosses.}
\label{bafe_eufe}
\end{figure}

The abundance ratio of barium, representative of the $s$-process, to europium
(representative of the $r$-process) is plotted in Fig.~\ref{bafe_eufe} as a
function of metallicity. For comparison we have also plotted [Ba/Eu] for the
sample of extremely metal-poor stars from Fran{\c c}ois et al.~(2007). 
Here SDSS J1036+1212 separates itself again from the sample of CEMP-r/s stars,
as the [Ba/Eu] ratio is more similar to that observed in EMP stars, where we
expect a pure $r$-process origin for the neutron-capture elements. 

\begin{figure}[ht]
\centering \includegraphics[width=0.5\textwidth]{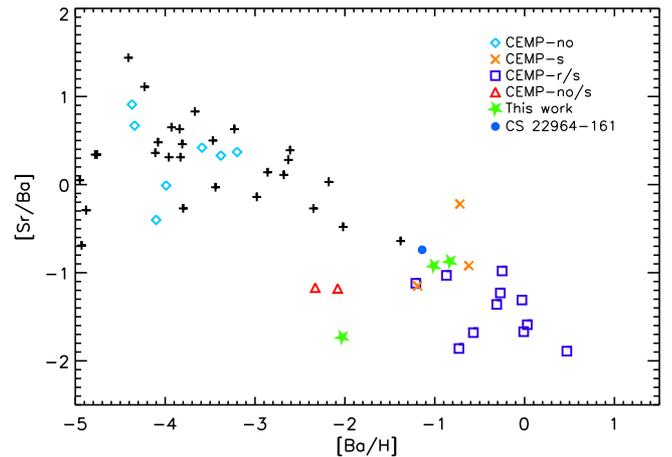}
\caption{The Sr/Ba ratio of our sample of stars compared to the classes of
  stars listed in Table~\ref{cemp_class} and the sample of extremely
  metal-poor stars from Fran{\c c}ois et al.~(2007).} 
\label{fig_srba}
\end{figure}

The [Ba/H] can be used as an indicator of the $s$-process efficiency of the
donor AGB star. We have plotted [Sr/Ba] as a function of [Ba/H] for the sample
of stars in Fig.~\ref{fig_srba}. A clear anti-correlation is seen between
[Sr/Ba] and [Ba/H] as well as a progression from the stars with less
$s$-process efficiency (CEMP-no, EMP) to those with high efficiency
(CEMP-r/s). The production ratio of heavy to light neutron-capture elements
increases with total production efficiency.

\begin{figure}[ht]
\centering \includegraphics[width=0.5\textwidth]{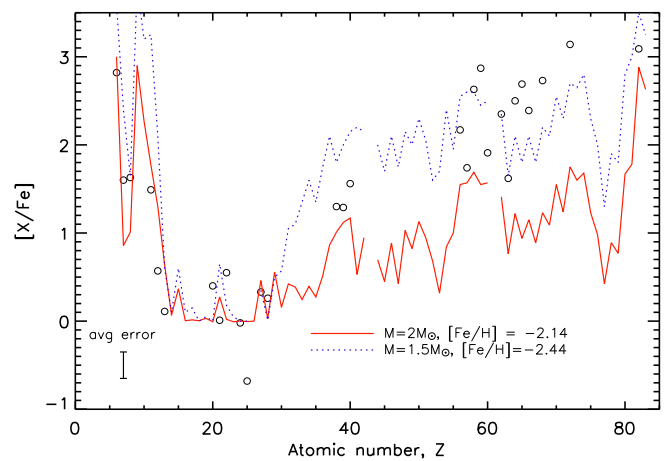}
\caption{Element distribution of SDSS J1349-0229 compared to model surface
  abundances of AGB stars. The first model, shown in red, has an initial mass
  of 2 M$_{\sun}$ and [Fe/H] = --2.14 (Cristallo et al.~2009). The second
  model has an initial mass of 1.5 M$_{\sun}$ and [Fe/H] = --2.44 (Cristallo
  et al.~2009b). }
\label{agb_A}
\end{figure}

\begin{figure}[ht]
\centering \includegraphics[width=0.5\textwidth]{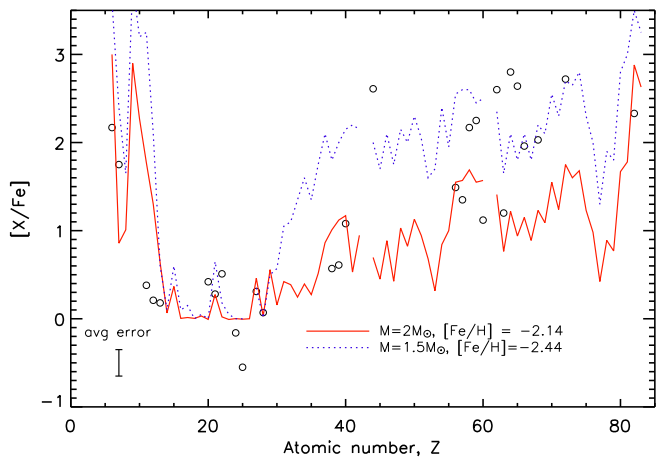}
\caption{As Fig.~\ref{agb_A} for SDSS J0912+0216.}
\label{agb_B}
\end{figure}

\begin{figure}[ht]
\centering \includegraphics[width=0.5\textwidth]{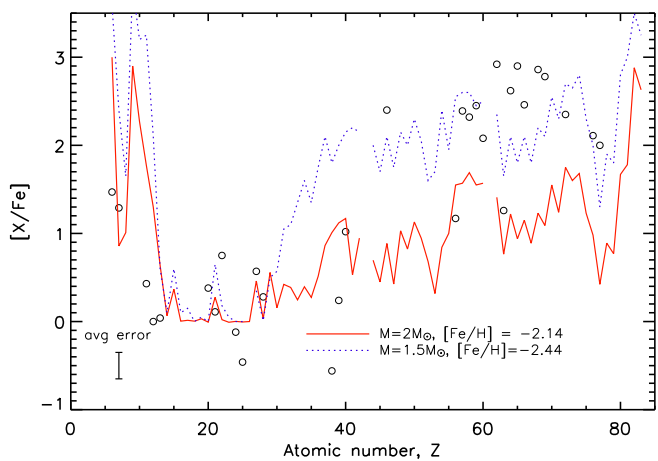}
\caption{As Fig.~\ref{agb_A} for SDSS J1036+1212.}
\label{agb_C}
\end{figure}

We compared the element distributions of our three stars to theoretical model surface
compositions of AGB stars. We employed models from \citet{cris2009a} and
\citet{cris2009b}. The results are shown in
Fig.~\ref{agb_A}, \ref{agb_B} and \ref{agb_C}. 
The models were computed using the FRANEC code (Chieffi et al.~1998). 
The first model used in the
comparison was computed with an initial mass of 2M$_{\sun}$ and [Fe/H] = --2.14. 
The second has an initial mass of 1.5M$_{\sun}$ and [Fe/H] = --2.44.

There is very little difference between the two models in the predicted
distribution of the light elements, but larger C, N and O values are
produced in the model with the lower mass and metallicity. The models diverge
at Z $>$ 30; the heavier elements are on average 1.0 dex more abundant in the
lower metallicity model compared to the high metallicity, high mass model. 

In general, the light $s$ elements (Sr, Y, Zr, hereafter $ls$) 
of the three stars are better reproduced by the
more massive AGB model, while the heavy $s$ elements (Ba to Hf, hereafter $hs$)
are better reproduced by the less massive
AGB model. We are however limited by the two models used in our
comparison. Bisterzo \& Gallino (2008) presented a comparison between a sample of 74
CEMP-s and CEMP-r/s stars and theoretical AGB models with different
$^{13}$C-pocket efficiencies, initial masses (M= 1.3,1.5,2,3 M$_{\sun}$),
and metallicities (--3 $<$ [Fe/H] $<$ --1).  
Their results showed that the low [$ls$/Fe] values found in CEMP stars can best
be reproduced at low metallicities with lower initial AGB masses ( M $<$ 1.4
M$_{\sun}$). Additionally, a strong initial $r-$enrichment of the order of [$r$/Fe] =
2.0 dex was needed to match the observations of several CEMP-r/s stars.

\section{Discussion and conclusions}

We classify SDSS J1349-0229 and SDSS J0912+0216 as CEMP-r+s stars. 
SDSS J1349-0229 shows clear radial-velocity variations following two sets of
observations, which indicates that it is a member of a binary system. SDSS
J1036+1212 shows similarities with CS 29528-041, a 
CEMP-no/s star. They have similar temperatures, gravities and
metallicities. They also both stand out as carbon-poor with respect
other CEMP stars. They differ in their nitrogen and $s$ and
$r$-process abundances however. The [$hs$/$ls$] ratio is higher for SDSS J1036+1212
compared to CS 29528-041, suggesting that the initial 
mass of the AGB companion star is lower for SDSS J1036+1212 than for CS 29528-041
(Bisterzo \& Gallino 2008). An AGB star with a higher initial mass would
produce less heavy neutron-capture elements (see Fig.~\ref{agb_A},
\ref{agb_B} and \ref{agb_C}). We measured [Ba/Fe] = 1.17 \relax in SDSS J1036+1212,
while the observed value in CS 29528-041 is 0.97. Furthermore, CS 29528-041 is
extremely enhanced in nitrogen: [N/Fe] = +3.07, while SDSS J1036+1212 is
moderately enhanced with [N/Fe] = +1.29. High initial mass AGB models by
\citet{herwig2004} with M $>$ 4M$\sun$ predict high nitrogen and low carbon,
whereas the lower mass models predict lower nitrogen and high carbon. The
high mass AGB models predict carbon abundances similar to those observed in
these two stars. A plausible origin can therefore be determined for CS
29528-041 in pollution from a high mass AGB star. The elements observed in SDSS
J1036+1212 point towards pollution by a low mass AGB companion. The low carbon
abundance though is puzzling and cannot be explained by current
models. Furthermore, the Li and [Ba/Eu] is similar to that observed in EMP
stars.

The scenario of mass transfer from an AGB companion is certainly
the one which allows the explanation of most of the chemical abundances
observed in the three stars studied in this work.
The high Li abundance observed in one of them poses some problems: if it is 
attributed to a production in the AGB star, it requires some
very fine tuning between Li production and destruction.
Finally it is somewhat surprising that all three stars
show a large (over 1\, dex) enhancement of Eu, which
is a pure r-process element.
This does not fit the AGB nucleosynthesis and requires
another origin. As all three stars show this
signature, it implies that such an event is relatively common
among CEMP stars.

\begin{acknowledgements}

We are grateful to R. Cayrel and
C.  Van't Veer for providing their
modified version of the BALMER code
and to P. Fran\c{c}ois for providing us 
the FITLINE code. 
We acknowledge financial support from  
EU contract MEXT-CT-2004-014265 (CIFIST).
We made use of the
supercomputing centre CINECA, which has granted us time to compute
part of the hydrodynamical models used in this investigation, through
the INAF-CINECA agreement 2006, 2007. 
This research has made use of the SIMBAD database, 
operated at CDS, Strasbourg, France, and of NASA's Astrophysiscs Data System.
Funding for the SDSS and SDSS-II has been provided by the Alfred P. Sloan Foundation, 
the Participating Institutions, the National Science Foundation, the U.S. Department of Energy, 
the National Aeronautics and Space Administration, the Japanese Monbukagakusho, 
the Max Planck Society, and the Higher Education Funding Council for England. 
The SDSS Web Site is http://www.sdss.org/.
The SDSS is managed by the Astrophysical Research Consortium for the Participating Institutions. 
The Participating Institutions are the American Museum of Natural History, Astrophysical Institute 
Potsdam, University of Basel, University of Cambridge, Case Western Reserve University, 
University of Chicago, Drexel University, Fermilab, the Institute for Advanced Study, 
the Japan Participation Group, Johns Hopkins University, the Joint Institute for Nuclear Astrophysics, 
the Kavli Institute for Particle Astrophysics and Cosmology, the Korean Scientist Group, 
the Chinese Academy of Sciences (LAMOST), Los Alamos National Laboratory, the 
Max-Planck-Institute for Astronomy (MPIA), the Max-Planck-Institute for Astrophysics (MPA), 
New Mexico State University, Ohio State University, University of Pittsburgh, University of 
Portsmouth, Princeton University, the United States Naval Observatory, and the University of 
Washington.

\end{acknowledgements}


\begin{thebibliography}{}

\bibitem[Adelman-McCarthy et al.(2007)]{dr5} 
Adelman-McCarthy, J.~K., et al.\ 2007, \apjs, 172, 634 

\bibitem[Adelman-McCarthy et al.(2008)]{dr6} 
Adelman-McCarthy, J.~K., et al.\ 2008, \apjs, 175, 297 

\bibitem[Aoki et al.(2002)]{aoki2002}Aoki W., Ryan S.G., Norris J.E., Beers
  T.C., Ando H., Tsangarides S., 2002, ApJ, 580, 1149 

\bibitem[Aoki et al.(2007)]{aoki2007} 
Aoki W., Beers T., Christlieb N., Norris J., Ryan S., \& Tsangarides S., 2007,
\apj, 655, 492

\bibitem[Aoki et al.(2008)]{aoki2008} 
Aoki W., Beers T., Sivarani T., Marsteller B., Lee Y.S., Honda S., Norris
J.E., Ryan S.G., Carollo D., 2008, \apj, 679, 1351

\bibitem[Asplund et al.(1999)] {asp99} Asplund, M., Nordlund, A., Trampedach,
  R., Stein, R., 1999, \aap, 346, L17

\bibitem[Asplund(2004)]{asp04} Asplund, M. 2004, MmSAI, 75, 300

\bibitem[Asplund \& Garc\'ia P\'erez(2001)]{asplund_OH2001} Asplund M., Garcia
  Perez A.E., 2001, \aap, 372, 601 

\bibitem[Barbuy et al.(1997)]{barbuy} Barbuy, B., Cayrel, R., Spite, M., 
Beers, T.~C., Spite, F., Nordstroem, B., \& Nissen, P.~E.\ 1997, \aap, 317, L63 


\bibitem[Barklem et 
al.(2000b)]{barklem00b} Barklem, P.~S., Piskunov, N., \& O'Mara, B.~J.\ 2000b, \aap, 363, 1091 

\bibitem[Barklem et 
al.(2000)]{barklem00} Barklem, P.~S., Piskunov, N., \& O'Mara, B.~J.\ 2000, \aap, 355, L5 


\bibitem[Baum\"uller \& Gehren(1997)]{baum1997} Baumueller D., Gehren T.,
  1997, \aap, 325, 1088 

\bibitem[Beers \& Christlieb(2005)]{BC2005} Beers, T.~C., \& Christlieb, N.\ 2005, \araa, 43, 531 

\bibitem[Behara et al.(2009)]{beharaJD10} Behara N.T., Ludwig H.-G., Bonifacio
  P., Sbordone L., Gonzalez Hernandez J.I., Caffau E., 2009, MmSAI, 80

\bibitem[Bergemann(2008)]{bergemann} Bergemann, M.\ 2008, Physica 
Scripta Volume T, 133, 014013 

\bibitem[Bergemann \& Gehren (2009)]{bergemann09} Bergemann, M.,Gehren T.,  \ 2009, 
IAU Symposium 265 eds. Cunha K., Spite, M. and ,Barbuy, B.  p. 

\bibitem[Bisterzo \& Gallino(2008)]{bg2008}
Bisterzo S., Gallino R., 2008, AIPC, 1001, 131	

\bibitem[Bonifacio \& Caffau(2003)]{BC03} Bonifacio, P., \& Caffau, E.\ 2003, \aap, 399, 1183 

\bibitem[Bonifacio et al.(2009a)]{bonifacioFS} Bonifacio P., Andersen J.,  Andrievsky S.M. 
et al.\ 2009a, in ``Science with VLT in the ELT era'' ed. A. Moorwood, Springer Verlag, Berlin,
p. 31 

\bibitem[Bonifacio et al.(2009b)]{B09}
Bonifacio, P., 	Spite M., Cayrel R., Hill V., Spite F., François P., Plez B.,
Ludwig H.-G., Caffau E., Molaro P., Depagne E., Andersen J., Barbuy B., Beers
T.C., Nordstr\"om B., Primas F., 2009b, \aap, 501, 519  

\bibitem[Bonifacio et al.(1998)]{bonifacio1998} Bonifacio, P., 
Molaro, P., Beers, T.~C., \& Vladilo, G.\ 1998, \aap, 332, 672 

\bibitem[Caffau et al. (2009)]{CLS} Caffau E.,  Ludwig H.-G., \& Steffen M. 2009, MmSAI, 80

\bibitem[Caffau \& Ludwig(2007)]{CL07} Caffau E., \& Ludwig H.-G., 2007, \aap, 467, L11 

\bibitem[Caffau et al.(2005)]{caffau05} Caffau, E., Bonifacio, P., 
Faraggiana, R., Fran{\c c}ois, P., Gratton, R.~G., \& Barbieri, M.\ 2005, \aap, 441, 533 

\bibitem[Caffau et al.(2008)]{caffau08} Caffau E., Sbordone L., Ludwig H.-G.,
Bonifacio P., Steffen M., Behara N.T., 2008, \aap, 483, 591 

\bibitem[Castelli(2005)]{atlas12} Castelli F., 2005a, MSAIS, 8, 25 

\bibitem[Castelli(2005)] {castelli2005} Castelli F., 2005b, MSAIS, 8, 44 

\bibitem[Castelli \& Kurucz(2003)]{CK03} Castelli, F., \& Kurucz, R.~L.\ 2003, 
in Modelling of Stellar Atmospheres, IAU Symp. No. 210,
eds. N. Piskunov et al., Poster A20, arXiv:astro-ph/0405087
 


\bibitem[Cayrel et al.(2004)]{cayrel04} Cayrel, R., et al.\ 2004, \aap, 416, 1117 

\bibitem[Chieffi et al.(1998)]{chieffi1998} Chieffi A., Limongi M., Straniero
  O., 1998, \apj, 502, 737

\bibitem[Christlieb et al.(2002)]{christlieb2002}
Christlieb N., Bessell M.S., Beers T.C., Gustafsson B., Korn A., Barklem P.S.,
Karlsson T., Mizuno-Wiedner M., Rossi S., 2002, Nature, 419, 904

\bibitem[Cohen et al.(2005)]{cohen05} Cohen, J.~G., et al.\ 
2005, \apjl, 633, L109 

\bibitem[Collet et al.(2007)]{collet2007} Collet, R., Asplund, M.,
  Trampedach, R., 2007, \aap, 469, 687

\bibitem[Cristallo et al.(2009a)]{cris2009a}
Cristallo S., Straniero O., Gallino R., Piersanti L., Dom\'inguez I., Lederer
M.T., 2009, \apj, 696, 797

\bibitem[Cristallo et al.(2009b)]{cris2009b}
Cristallo S., Piersanti L., Straniero O., Gallino R., Dominguez I., Kappeler
F., 2009, PASA, arXiv:astro-ph/0904.4173 

\bibitem[Dekker et al.(2000)]{dekker} Dekker, H., D'Odorico, 
S., Kaufer, A., Delabre, B., \& Kotzlowski, H., 2000, \procspie, 4008, 534 

\bibitem[Den Hartog et al.(2003)]{den2003}Den Hartog, E.A., Lawler, J.E., Sneden, C., Cowan, J.J., 2003, ApJSS, 148, 543

\bibitem[Den Hartog et al.(2006)]{den2006}Den Hartog E.A., Lawler J.E., Sneden
  C., Cowan J.J., 2006 ApJSS 167, 292

\bibitem[Drawin(1969)]{Drawin} Drawin H.W., 1969, Z. Physik 225, 483

\bibitem[Fran\c{c}ois et al.(2003)]{fdv03} 
 Fran\c{c}ois P., Depagne E., Hill V. et al., 2003,  A\&A, 403, 1105

\bibitem[Fran\c{c}ois et al.(2007)]{fdv07} 
Fran\c{c}ois P., Depagne E., Hill V. et al., 2007, \aap, 476, 935

\bibitem[Frebel et al.(2005)]{frebel2005}
Frebel A., Aoki W., Christlieb N. et al., 2005, Nature, 434, 871

\bibitem[Freytag et al.(2002)]{fsd02} 
 Freytag B., Steffen M., Dorch B., AN, 323, 213

\bibitem[Goldman \& Gillis(1981)]{gg1981} Goldman A., \& Gillis J.~R., 1981, JQSRT,  25, 111 


\bibitem[Gonz\'alez Hern\'andez et al.(2008)]{jonay}
Gonz\'alez Hern\'andez, J., Bonifacio, P., Ludwig, H.-G., et al.\ 
2008, A\&A, 480, 233 

\bibitem[Gratton et al.(1999)]{gratton1999}Gratton R.G., Carretta E., Eriksson
  K., \& Gustafsson B., 1999, \aap, 350, 955 

\bibitem[Herwig(2004)]{herwig2004} Herwig F., 2004, ApJS, 155, 651

\bibitem[Hill et al.(2000)]{hill} Hill, V., Barbuy B., Spite M., Spite F., Cayrel R., Plez B., Beers T.C.,
Nordstr\"om B., Nissen P.E., 2000, \aap, 353, 557 

\bibitem[Ivarsson et al.(2001)]{ivar2001}Ivarsson S., Litz\'en U., Wahlgren G.M., 2001, Phy.Scr., 64, 455	

\bibitem[Johnson \& Bolte(2004)]{JohnsonBolte}Johnson J.A., \& Bolte M., 2004, ApJ, 605, 462

\bibitem[Jonsell et al.(2006)]{jonsell}
Jonsell K., Barklem P.S., Gustafsson B., Christlieb, N., Hill V., Beers T.C.,
Holmberg J., 2006, \aap, 451, 651

\bibitem[Karakas \& Lattanzio(2007)]{kl2007} Karakas A.I., Lattanzio J.C., 2007, PASA, 24, 103

\bibitem[Kurucz(1993)]{KCD13} Kurucz R., 1993, ATLAS9 
Stellar Atmosphere Programs and 2 km/s grid.~Kurucz CD-ROM No.~13.~ 
Cambridge, Mass.: Smithsonian Astrophysical Observatory, 1993., 13,

\bibitem[Kurucz(1996)]{kurucz96} Kurucz R.L., 1996, in IAU Symp. 176, ed. K.G. Strassmeier
  \& J.L. Linsky (Dordrecht: Kluwer), 523  

\bibitem[Kurucz(2005a)]{K05} Kurucz R.~L., 2005a, MSAIS, 
 8, 14 

\bibitem[Kurucz(2005b)]{Kweb} Kurucz R.L., 2005b, kurucz.harvard.edu/ 

\bibitem[Lawler et al.(2001a)]{lawler_la2}Lawler J.E., Bonvallet G., Sneden
  C., 2001a, ApJ 556, 452 

\bibitem[Lawler et al.(2001b)]{lawler2001}Lawler J.E., Wickliffe M.E., Cowley
  C.R., Sneden C., 2001b, ApJSS, 137, 341

\bibitem[Lawler et al.(2006)]{lawler2006}Lawler J.E., Den Hartog E.A., Sneden
  C., Cowan J.J., 2006, ApJSS 162, 227 

\bibitem[Lawler et al.(2007)]{law07} Lawler J.E., Den Hartog E.A., Labby Z.E.,
  Sneden C., Cowan J.J., Ivans I.I., 2007, \apjs, 169, 120

\bibitem[Lawler et al.(2008)]{law08} Lawler J.E., Sneden C., 
Cowan J.J., Wyart J.-F., Ivans I.I., Sobeck J.S.,
Stockett M.H., \& Den Hartog E.A. 2008, \apjs, 178, 71

\bibitem[Lawler et al.(2009)]{lawler2009} Lawler J.E., Sneden C., Cowan J.J.,
  Ivans I.I., Den Hartog E.A., 2009, ApJS, 182, 51

\bibitem[Li et al.(2007)]{li2007}Li R., Chatelain R., Holt R.A., Rehse S.J.,
  Rosner S.D., Scholl T.J., 2007, Phys.Scr., 76, 577 


\bibitem[Lodders(2003)]{lodders} Lodders, K.\ 2003, \apj, 591, 1220 

\bibitem[Lucatello et al.(2005)]{lucatello2005}
Lucatello S, Tsangarides S., Beers T.C., Carretta E., Gratton R.G., Ryan S.G.,
2005, \apj, 625, 825


\bibitem[Ludwig et al.(2009)]{ludwigjd10} Ludwig, H.-G., Caffau, E., 
Steffen, M., Freytag, B., Bonifacio, P.\ 2009, MmSAI, 80, 

\bibitem[Lucatello et al.(2006)]{lucatello06} Lucatello, S., Beers, 
T.~C., Christlieb, N., Barklem, P.~S., Rossi, S., Marsteller, B., Sivarani, 
T., \& Lee, Y.~S.\ 2006, \apjl, 652, L37 

\bibitem[Marigo(2001)]{marigo} Marigo, P.\ 2001, \aap, 370, 194 

\bibitem[McWilliam(1998)]{mcwill1998}
McWilliam, A., 1998, AJ, 115, 1640


\bibitem[Mihalas(1978)]{mihalas} Mihalas, D.\ 1978, 
``Stellar atmospheres''
San 
Francisco, W.~H.~Freeman and Co., 1978.~650 p.,

\bibitem[Norris et al.(1997)]{norris1997} Norris J.~E., Ryan S.~G., \& Beers
  T.~C.\ 1997b, \apj, 488, 350  

\bibitem[Norris et al.(2007)]{norris2007}
Norris J.E., Christlieb N., Korn A.J., Eriksson K., Bessell M.S., Beers T.C.,
Wisotzki L., Reimers D., 2007, \apj, 670, 774

\bibitem[Paunzen et al.(1999)]{paunzen} Paunzen E., Kamp I., Iliev I.Kh.,
  Heiter U., Hempel M., Weiss W.W., Barzova I.S., Kerber F., Mittermayer P., 1999, \aap, 345, 597 

\bibitem[Reyniers et al.(2002)]{reyniers2002}
Reyniers M., Van Winckel H., Bi\'emont E., Quinet P., 2002, \aap, 395, 35

\bibitem[Ryan(1998)]{Ryan1998} Ryan, S.~G.\ 1998, \aap, 331, 
1051

\bibitem[Sbordone et al.(2004)]{luca2004} Sbordone L.,  Bonifacio P., Castelli
  F., Kurucz R.L., 2004, MSAIS, 5, 93

\bibitem[Sbordone(2005)]{luca2005} Sbordone L., 2005, MSAIS, 8, 61 

\bibitem[Sivarani et al.(2006)]{sivarani}
Sivarani T., Beers T.C., Bonifacio P., et al., 2006, A\&A 459, 125 

\bibitem[Sneden et al.(2003)]{sneden2003}Sneden C., Cowan J.J., Lawler J.E.,
  et al., 2003, ApJ, 591, 936


\bibitem[Spite \& Spite(1982a)]{spite82Natur} Spite, M., \& Spite, F.\ 1982a, \nat, 297, 483 

\bibitem[Spite \& Spite(1982b)]{spite82} Spite, F., \& Spite, M.\ 1982b, \aap, 115, 357 


\bibitem[Spite et al.(2005)]{spite2005}	Spite M., Cayrel R., Plez B. et al.,
  2005, \aap, 430, 655 

\bibitem[Stancliffe(2009)]{Stancliffe} Stancliffe R.J. 2009, MNRAS,
  394, 1051

\bibitem[Steenbock \& Holweger(1984)]{SH} Steenbock, W., 
\& Holweger, H.\ 1984, \aap, 130, 319 


\bibitem[Stehl{\'e} 
\& Hutcheon(1999)]{stehle} Stehl{\'e}, C., \& Hutcheon, R.\ 1999, \aaps, 140, 93 

\bibitem[St{\"u}renburg \& Holweger(1990)]{stuerenburg} St{\"u}renburg, S., \& Holweger, H.\ 1990, \aap, 237, 125

\bibitem[Thompson et al.(2008)]{thompson2008}
Thompson I.B., Ivans I.I., Bisterzo S., Sneden C., Gallino R., Vauclair S.,
Burley G.S., Shectman S.A., Preston G.W., 2008, \apj, 677, 556


\bibitem[Ventura \& Marigo(2009)]{VM09} Ventura, P., \& Marigo, P.\ 2009, MNRAS in press,
arXiv:0907.3204 


\bibitem[Ventura et al.(2002)]{ventura} Ventura, P., D'Antona, F., \& Mazzitelli, I.\ 2002, \aap, 393, 215

\bibitem[York et al.(2000)]{york} York D.~G., Adelman J., Anderson J.E. et al., 2000, \aj, 120, 1579 

\bibitem[Wahlgren(2005)]{wahl2005}Wahlgren G.M., 2005, MSAIS, 8, 108

\bibitem[Wedemeyer et al(2004)]{wfs04} 
 Wedemeyer S., Freytag B., Steffen M., Ludwig H.-G., Holweger H., \aap, 414, 1121 

\bibitem[Wickliffe \& Lawler(1997)]{wick1997}Wickliffe M.E., \& Lawler J.E., 1997 JOSAB, 14, 737

\bibitem[Wickliffe et al.(2000)]{wick2000}Wickliffe M.E., Lawler J.E., Nave
  G., 2000, JQRST, 66, 363

\end{thebibliography}
\end{document}